\documentclass[3p,twocolumn]{elsarticle}

\usepackage{lineno,hyperref}

\usepackage{amsmath}
\usepackage{amssymb}
\usepackage{amsbsy}
\usepackage{makeidx}
\usepackage{epsf}
\usepackage{graphics}
\usepackage{graphicx}
\usepackage{amsfonts}
\usepackage{subfigure}
\usepackage{color}
\usepackage[update,prepend]{epstopdf}

\journal{Physics Letters A}

\begin{document}

\begin{frontmatter}

\title{Black holes will break up solitons and white holes may destroy them}

\author{Fiki T.\ Akbar} 
\ead{ftakbar@fi.itb.ac.id}
\author{Bobby E.\ Gunara}
\address{Theoretical Physics Laboratory, Theoretical High Energy Physics and Instrumentation Research Group,\\ Faculty of Mathematics and Natural Sciences, Institut Teknologi Bandung, Jl.\ Ganesha no.\ 10 Bandung, 40132, Indonesia}
\ead{bobby@fi.itb.ac.id}
\author{Hadi Susanto\corref{cor1}}
\address{Department of Mathematical Sciences, University of Essex, Colchester, CO4~3SQ, United Kingdom}
\cortext[cor1]{Corresponding author}
\ead{hsusanto@essex.ac.uk}

\begin{abstract}
We consider a quantum analogue of black holes and white holes using Bose-Einstein condensates. The model is described by the nonlinear Schr\"odinger equation with a 'stream flow' potential, that induces a spatial translation to standing waves. We then mainly consider the dynamics of dark solitons in a black hole or white hole flow analogue and their interactions with the event horizon. A reduced equation describing the position of the dark solitons was obtained using variational method. Through numerical computations and comparisons with the analytical approximation we show that solitons can pass through black hole horizons even though they will break up into several solitons after the collision. In the interaction with a white hole horizon, we show that solitons either pass through the horizon or will be destroyed by it. 
\end{abstract}


\end{frontmatter}


\section{Introduction}

The Einstein field equations from the theory of general relativity describe the fundamental interaction of gravitation as a result of space-time being curved by matter and energy {densities} \cite{eins15,eins16}. All metric modelling a physical system must satisfy the field equations. A special solution for the equations under some assumptions was obtained by Schwarzschild \cite{schw16}. It was later understood that the solution describes a black hole \cite{fink58,krus60}, i.e.\ a region of space with so much concentrated mass that a nearby object, including light, cannot escape its gravitational pull. {The solution was later generalized to the charged black holes which is a static solution of Einstein-Maxwell equation known as The Reissner-Nordstr\"{o}m black holes and the rotating black holes known as Kerr black holes \cite{kerr63}}. 

Black hole candidates have been identified, e.g., in \cite{webs72,bolt72} through electromagnetic observations by detecting their effect on other matter nearby (see the review \cite{casa14}). The first direct observation of a binary black hole system merging to form a single black hole was reported in \cite{abbo16,abbo16b} through the detection of the so-called gravitational waves \cite{eins16b,eins18}. However, despite the successful detections of black holes, experiments with gravitational black holes in general relativity will not happen soon. 
It is because generating {a black hole requires a sophisticated method to handle its extreme strong gravitational fields}, which is beyond the current technology. Noting that  {the geometry near}  black holes are like {spacetime}  rivers, the difficulty is overcome by turning to simulating aspects of general relativity through black hole analogues made of physical systems possessing an "event horizon" \cite{nove02,schu07,barc11,robe12}. Horizons are "the point of no return", that in the gravitational fields are defined as the point at which the gravitational pull becomes so great as to make escape impossible. 

The idea of black hole analogues was first proposed by Unruh \cite{unru81} in fluid flow. It is commonly pictured as a river flowing \emph{towards} a waterfall with the flow speed increasing in the same direction of the flow \cite{hami08}. If the river is populated by fish with maximum speed $c$, then the event horizon is the point where the river flow speed is also $c$ beyond which the fish can no longer swim upstream. 
Additionally in this analogue one can also have a white hole, which is modelled by a river flowing \emph{from} a waterfall with speed flow that was initially very fast at the waterfall and slows down as it travels farther. The event horizon is still defined in the same way, but the populated fish, instead of not being able to leave the region beyond the horizon, in here cannot enter it. 
Experiments using this analogue have been reported in \cite{rous08,wein11}.

A black hole analogue also has been proposed using Bose-Einstein condensates \cite{gara00,gara01,barc01,barc01b}. Experiments with a transonic flowing ultracold atomic condensate have been reported as well \cite{laha10,stei14}. The system that is modelled by the nonlinear Schr\"odinger equation with potential, i.e.\ the Gross-Pitaevskii equation. The subsonic and the supersonic region is obtained by introducing inhomogeneously piecewise constant potentials admitting homogeneous solutions. Stationary solutions and their stability were analysed in \cite{mich13}. Time dynamics of the transonic flow was studied in \cite{mich15}. It was observed that white hole flows typically yield undulations in the supersonic region and soliton trains in the subsonic one, while black holes exhibit a correspondence with the so-called no-hair theorem in gravitational black holes \cite{mich16}.

It is important to note that { in Bose-Einstein condensates the  prodution of solitons from unstable white flows \cite{mich15} is indeed due to nonlinearity.  } As it is known that defocusing Bose-Einstein condensates admit dark solitons (see the review \cite{fran10,kevr15}), here we aim to study systematically the interaction between dark solitons and the event horizon in both black hole and white hole flows in Bose-Einstein condensates (as opposed to the creation of dark solitons reported previously). 

We show that dark solitons can pass black holes. However, the flows always break up the solitons into several ones. In the interaction with white holes, dark solitons may either pass the event horizon or be destroyed by it yielding shock waves or undulations \cite{kamc00,el16} both in the subsonic and supersonic region. Based on a variational formulation, we derive an effective equation describing the soliton position, whose zeros from comparisons with the numerics give a sufficient condition for the destruction. We also discuss the applicability of the effective equation, where we obtain that the analysis is only valid when the relative speed between the subsonic and supersonic region is not too large. 

\section{Mathematical model}

\begin{figure*}[tbhp!]
\centering
{
\subfigure{\includegraphics[width=7cm]{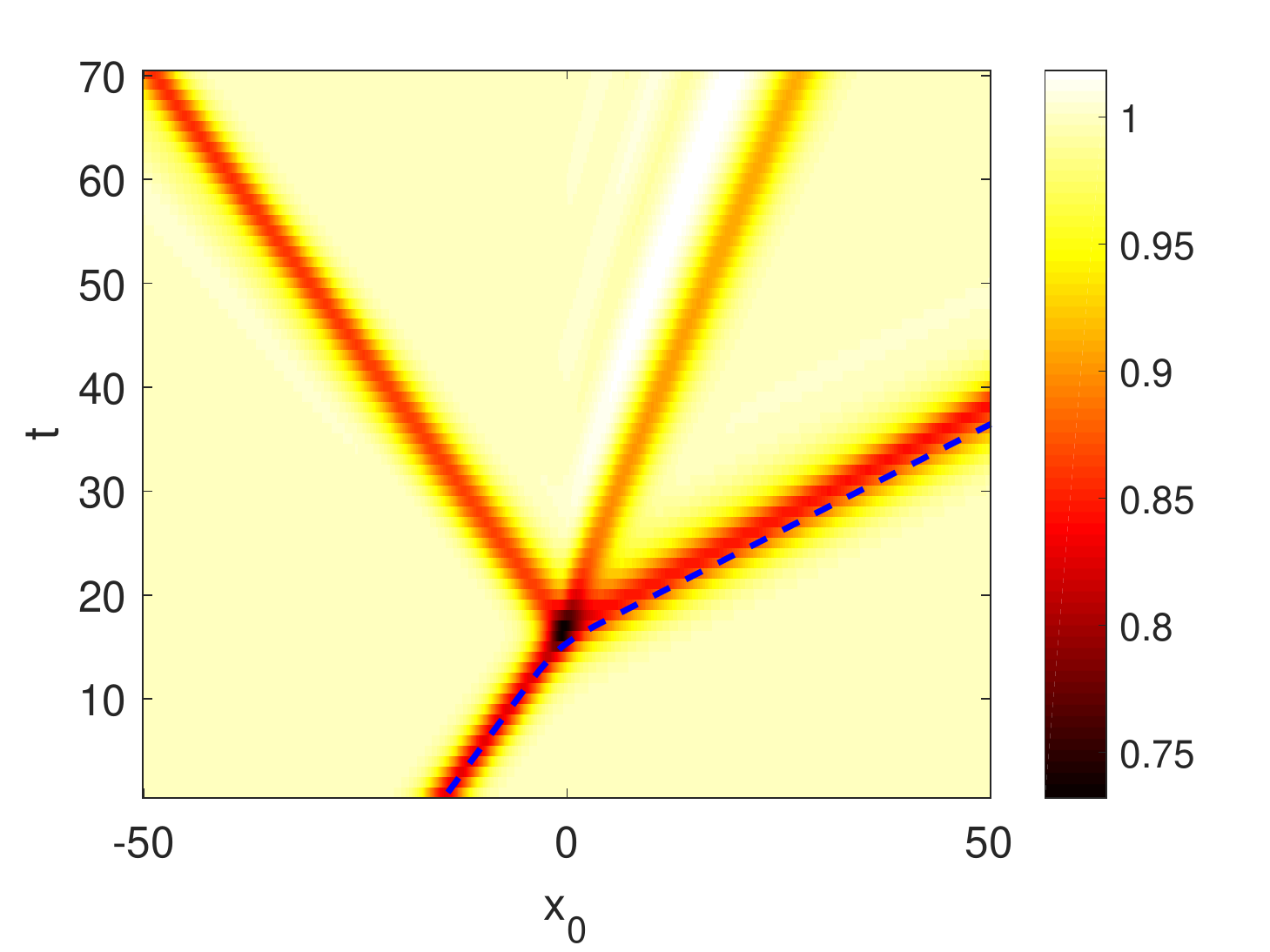}}
\subfigure{\includegraphics[width=7cm]{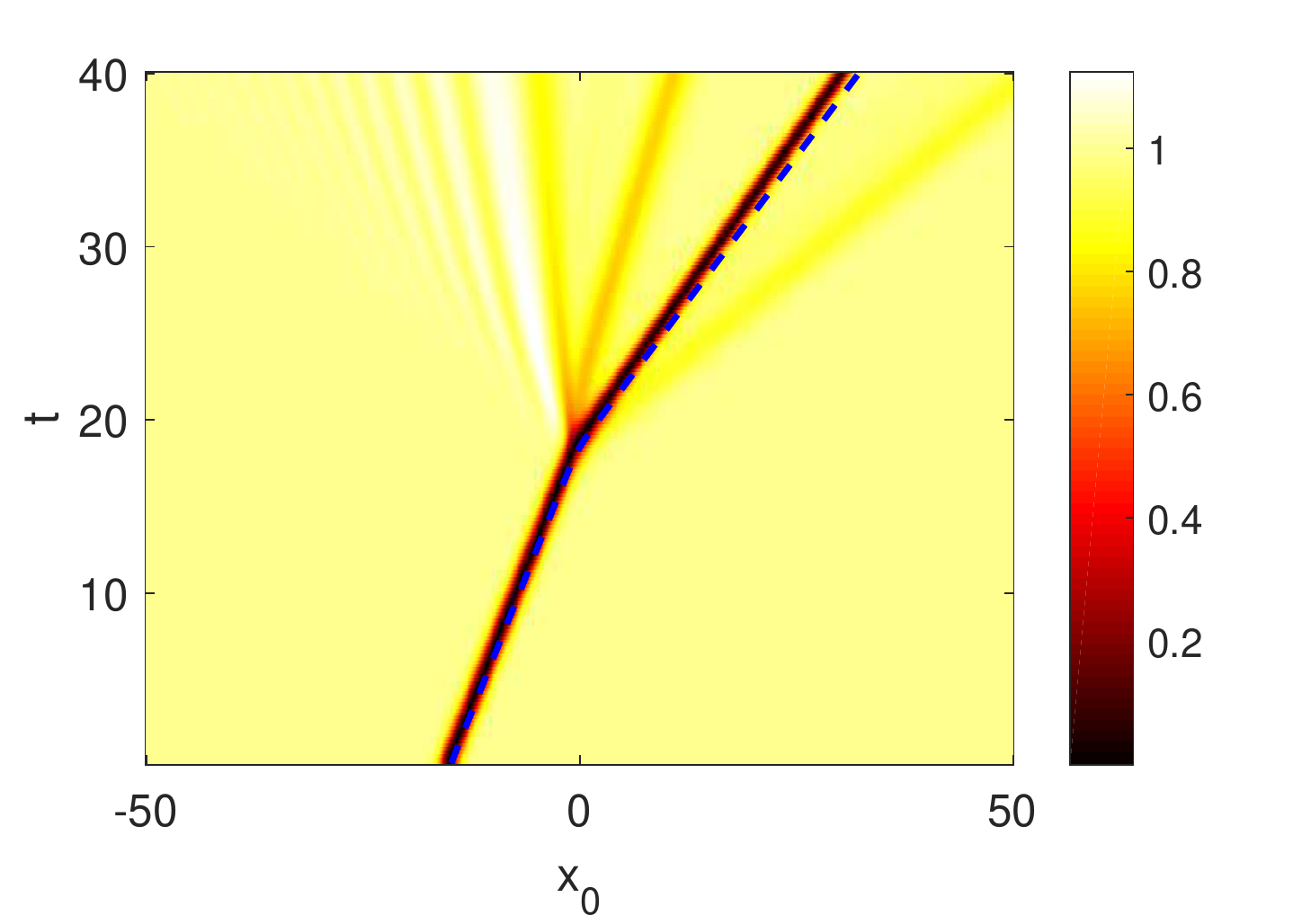} }
}
\caption{The top view of the dynamics of a dark soliton when interacting with the event horizon of a black hole for $v_r=1.5$ and (top) $v_l=0,\,v=0.9$, (bottom) $v_l=0.8,\,v=0$. The dashed lines are from the approximation \eqref{pos}.}
\label{fig1} 
\end{figure*}

We consider the defocusing nonlinear Schr\"odinger equation of the form
\begin{equation}
iu_t=-i  {{\mathcal V}}(x)u_x-\frac12u_{xx}-\omega u+g|u|^2u,
\label{gov}
\end{equation}
with the propagation constant $\omega$, the repulsive interaction coefficient $g$ and the 'stream' potential representing the local velocity
\begin{equation}
{{\mathcal V}}(x) = 
\left\{\begin{array}{lll}
v_l, &x\leq0,\\
v_r, &x>0.
\end{array}
\right.
\label{pot}
\end{equation}
In previous works, motivated by the experiments \cite{laha10,stei14} the black/white hole analogues were considered using ${{\mathcal V}}\equiv0$, $\omega=\omega(x)$, $g=g(x)$ 
\cite{mich13,mich15,mich16}. The spatially dependent $\omega$ and $g$ provide the "waterfall'' potential. Here, we consider constant $\omega$ and $g$ and use ${{\mathcal V}}(x)$ instead to yield the "waterfall'' potential. Without loss of generality, we take $\omega=g=1$ and hence the speed of sound $c=1$ \cite{abid03}, as such that the black and white hole will correspond to the case of $0<v_l<1<v_r$ and $0<v_r<1<v_l$, respectively. 

Our theoretical setup allows us to view the interaction dynamics between solitons and a black/white hole as caused solely by the stream velocity change. Using inhomogeneous $\omega$ and $g$ may create dynamics that is present due to the spatial dependence of the linear potential $\omega$ and nonlinear interaction $g$, such as in the study of the so-called collisionally inhomogeneous Bose-Einstein condensates and nonautonomous nonlinar optics (see, e.g., \cite{mara10} and references therein). Our model has the same steady state properties as those of the model considered in \cite{mich13,mich15,mich16}, i.e.\ the state remains transonic and becomes stationary $u_t\to0$ as $t\to\infty$.

The governing equation \eqref{gov} can be written in a hydrodynamic representation. Using the Madelung transformation $u(x,t) = R(x,t)e^{{i}\theta(t,x)}$ and substituting it into \eqref{gov} will yield 
\begin{align}
&R_t= -R_x\theta_{x} -\frac12R\theta_{xx} - \mathcal{V}(x)R_{x},\label{kontinuitas}\\
&R(\partial_t+{\mathcal V}(x)\partial_x)\theta = \frac12\left(\partial_{xx}-\theta_x^2+2-2R^2\right)R.\label{p2}
\end{align}
Equation \eqref{kontinuitas} can be interpreted as the continuity equation (i.e.\ the fluid conservation equation) for the density $R^2$ and the phase $\theta_{x}$. The steady-state solutions $R_t=\theta_t=0$ can be analysed by integrating \eqref{kontinuitas} to obtain 
\[
\theta_x=-\mathcal{V}(x)\left(1-\frac{1}{R^2}\right),
\]
where we have assumed that $\mathcal{V}(x)$ is piecewise-constant and as $x\to\pm\infty$, $R^2\to1$ and $\phi_x\to0$. Substituting it into the time-independent equation of \eqref{p2} will give 
\begin{equation}
R_{xx}=-\mathcal{V}(x)\left(R-\frac1{R^3}\right)-2(1-R^2)R.
\label{pp}
\end{equation}
The phase-plane analysis of \eqref{pp} is simple through constructing a composite phase-portrait \cite{mara10}. One can check that in the subsonic and supersonic regime, $R=1$ is respectively a saddle point and a centre, i.e.\ the same as the observation in \cite{mich13}. 

When ${{\mathcal V}}(x)\equiv V=\texttt{const.}$, Eq.\ \eqref{gov} has a dark soliton solution of the form
\begin{equation}
u(x,t)=A\tanh A\left(x-x_0\right)+iv,
\label{ds}
\end{equation}
with $x_0=(v+V)t$ and $A^2+v^2=1$. The soliton is black when $v=0$. It is clear that $V$ acts as a stream flow that drifts the soliton with relative velocity $v+V$. One can also note that for any $|V|<1$, there is a dark soliton with "internal" velocity $|v|>|V|$, such that the soliton is travelling in the opposite direction to the stream flow. When $|V|>1$, there is no longer standing soliton. The soliton is therefore like a 'quantum fish' swimming in a 'quantum waterfall'.  

In the following we will study the influence of the even horizon that is located at $x=0$ to the soliton. Under the inhomogeneous potential \eqref{pot}, 
one can study analytically the dynamics of the dark soliton adiabatically using variational methods as in, e.g., \cite{kivs95}. Using the ansatz \eqref{ds} where $x_0$ is now unknown and a function of $t$, the position of the dark soliton is described by
\begin{equation}
{x_0}_t=v+\frac{v_l+v_r}2-\frac{v_l-v_r}2\tanh[\sqrt{1-v^2}x_0].
\label{pos}
\end{equation}

Equation \eqref{pos} has one and one only equilibrium given by ${x_0}_t=0$ that exists if the inequality 
\begin{equation}
-1<\frac{2v+{v_l+v_r}}{{v_l-v_r}}<1
\label{ineq}
\end{equation}
is satisfied. In this case, a moving soliton may stop. The stability of the equilibrium is clearly either an attractor or a repeller. 

In the following we will solve the governing equations numerically and compare it with the analytical approximation above. 

\section{Discussion}
\begin{figure}[tbhp]
\centering
{
\subfigure{\includegraphics[width=7cm]{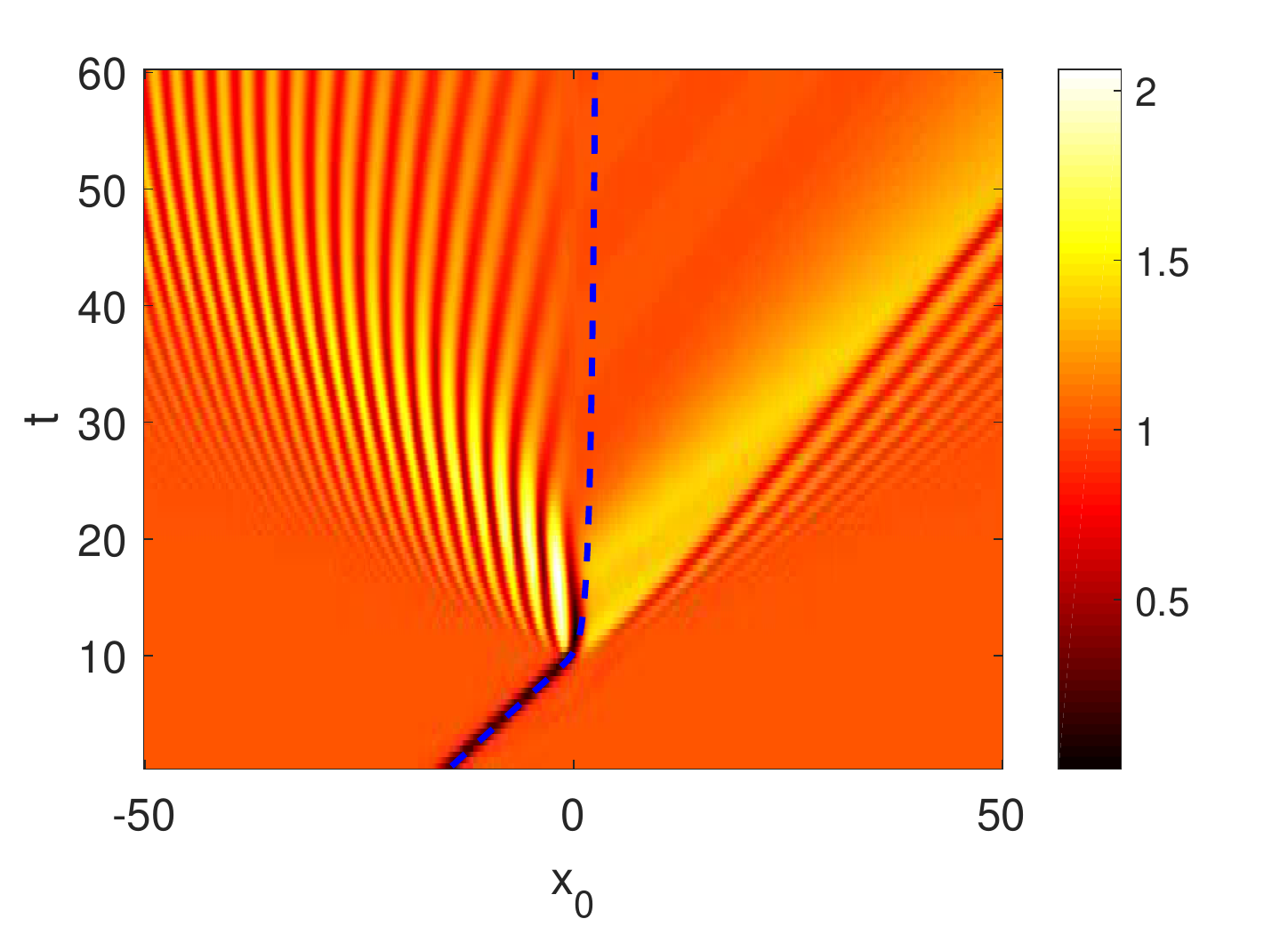}}
\subfigure{\includegraphics[width=7cm]{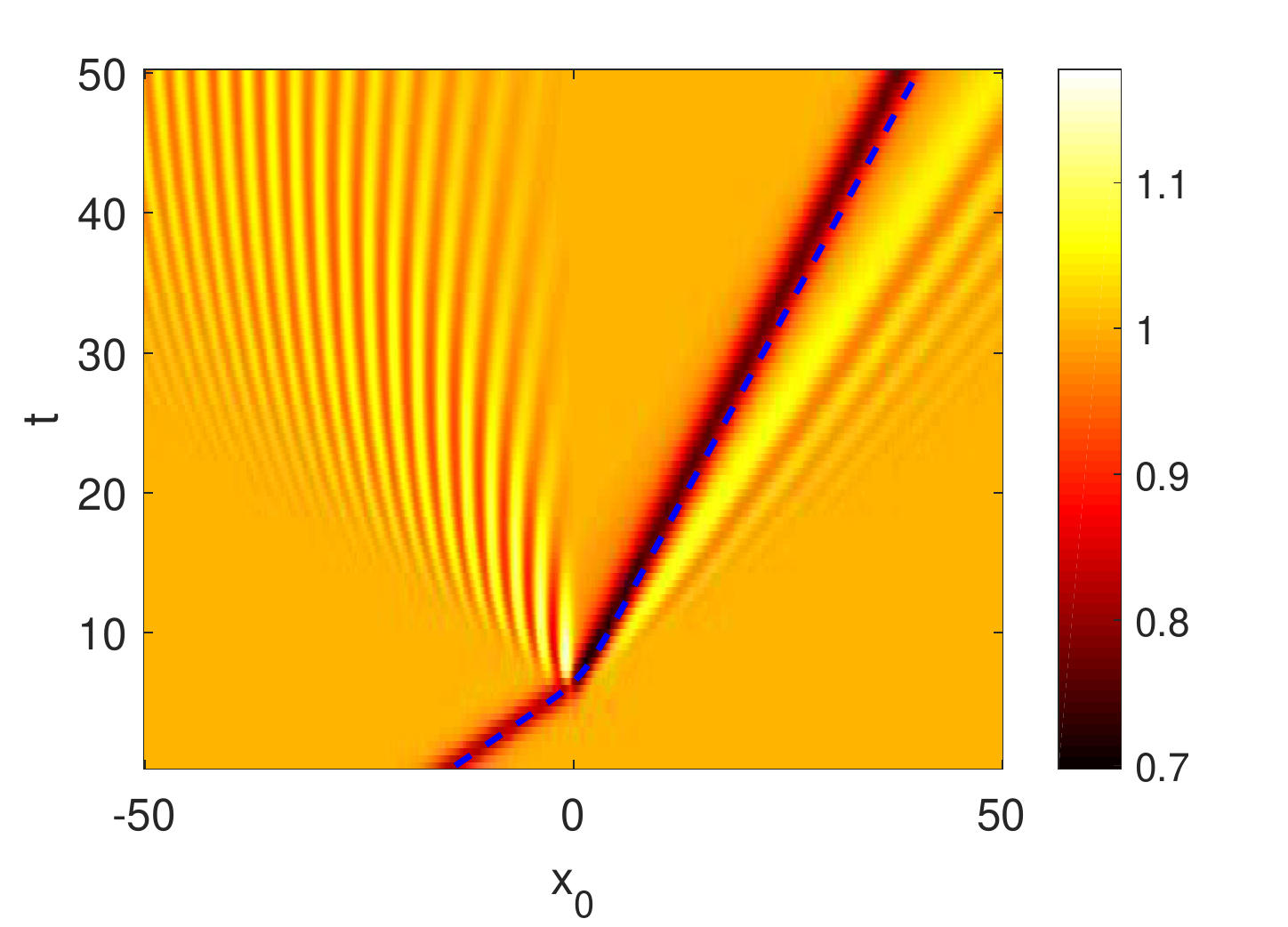} }
}
\caption{The same as in Fig.\ \ref{fig1}, but for white holes with $v_l=1.5,\,v_r=0,$ and (top) $v=-0.1$, (bottom) $v=0.9$.}
\label{fig3} 
\end{figure}

First, we consider black hole configurations. Shown in Fig.\ \ref{fig1} is the top view of two time dynamics of a dark  soliton that is incoming towards the event horizon at $x=0$. In both cases, in the supersonic regime $v_r=1.5$. For the first simulation, we choose the condition that the fluid in the region $x<0$ does not flow, i.e.\ $v_l=0$. The dark soliton is instead travelling with velocity $v=0.9$ towards the origin. Due to the collision with the event horizon, one can observe that the soliton breaks up into three dark solitons after impact. Shock wave is also generated in the supersonic regime behind the middle soliton. However, the wave cannot cross the event horizon. Additionally we observe that the background $|\psi|=1$ behaves as an 'attractor', that pushes all the disturbances away towards $x\to\pm\infty$, which is in agreement with the 'no-hair theorem' analogue reported in \cite{mich16}.


We also plot the approximation \eqref{pos} as dashed line, where one can see that it approximates well the position of one of the three dark solitons. 

Next, we consider the dynamics of a dark soliton that is moved by the stream in the subsonic regime towards the event horizon. Shown in Fig.\ \ref{fig1}(b) is the case when the internal velocity $v$ vanishes (i.e.\ the soliton is black), but the flow in the subsonic regime is $v_l=0.8$. Similarly to panel (a), we obtain three solitons after collision. To be precise the collision yields one dark and two grey solitons. The variational approximation \eqref{pos} also shows the position of the dark soliton. However, there is a difference in that this time the shock wave is generated in the subsonic regime and is travelling towards $x\to-\infty$.

For black holes, i.e.\ $0\leq v_l<1<v_r$, the inequality \eqref{ineq} cannot be satisfied. Therefore, we conjecture that dark solitons will always pass the event horizon, but will break up into several solitons. This is confirmed in all our numerical simulations. 

Next, we consider the dynamics of grey or dark solitons in white holes. We show in Fig.\ \ref{fig3} the dynamics of a dark soliton when it interacts with the event horizon.  In panel (a), the dark soliton is moving against the supersonic flow, i.e.\ the internal velocity $-1<v<0$. As there is no soliton that can escape the flow, the soliton later collides with the event horizon. The outcome of the collision is the destruction of the soliton and the generation of shock waves both in the supersonic and subsonic regimes. The approximation is also shown in dashed line, which is in agreement with the numerics of the full equation. The inequality \eqref{ineq} is indeed satisfied in this case.

We also show the case when the condition \eqref{ineq} is not met. Panel (b) depicts the dynamic of the soliton passing the event horizon, which is in agreement with the theoretical prediction. The generated shock waves can be seen also existing both in the supersonic and subsonic regimes. 

\begin{figure}
\centering
{
{\includegraphics[width=7cm]{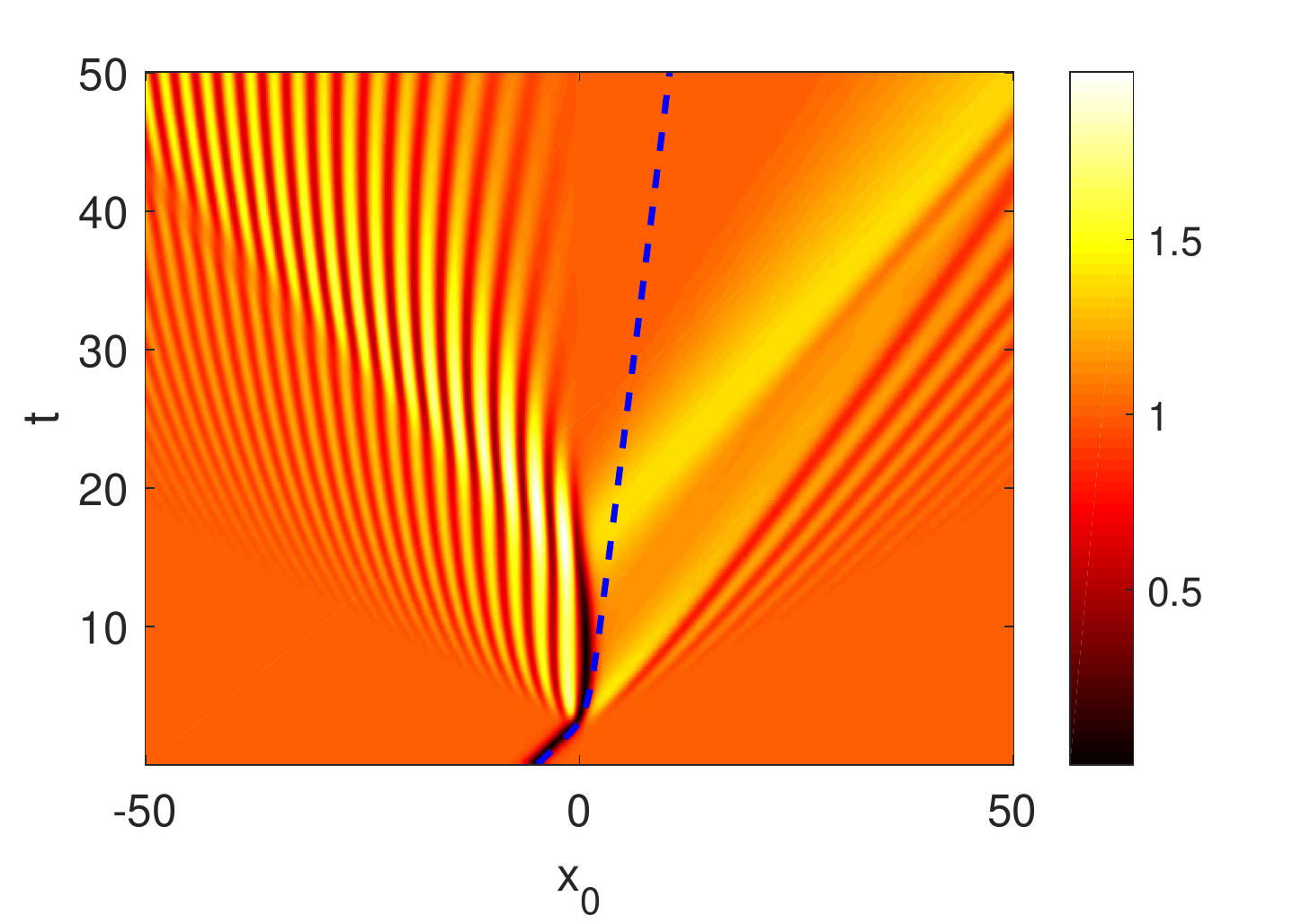}}
}
\caption{The same as Fig.\ \ref{fig3}, but with $v=0.2$.}
\label{fig4} 
\end{figure}

Despite the agreement between the numerics and the approximation \eqref{pos} we obtained in Figs.\ \ref{fig1}-\ref{fig3}, we also saw discrepancies as shown in Fig.\ \ref{fig4}. While the condition for an equilibrium is violated, the numerics of the original full equation still shows a destruction of the incoming soliton. The actual critical velocity derived from the full equation that is needed to cross the horizon is larger than that predicted by \eqref{ineq}. It is rather expected because \eqref{pos} was derived without considering the presence of radiation upon interaction. This shows that \eqref{ineq} rather provides a sufficient, but not necessary, condition for soliton destructions. 

\section{Conclusion}

We have considered the dynamics of dark solitons in a black hole or white hole flow and their interactions with the event horizon. We showed numerically and analytically that solitons can pass through black hole horizons even though they will break up into several solitons after the collision. In the interaction with a white hole horizon, typically solitons will either pass through the horizon or will be destroyed by it. A sufficient condition for the destruction was derived through the equation describing the position of the dark solitons, that was obtained using a variational method. The method presented herein can be applied readily to the theoretical setup of \cite{mich13,mich15,mich16} by including the variation of $\omega$ and $g$ in \eqref{pos}.

A natural follow-up to the work here 
would be to see if the typical dynamics reported herein is also observed in different physical models. The interaction of non-topological solitons, i.e.\ such as breathers, of a nonlinear wave equation with event horizons (note that Unruh's interesting proposal for black hole analogues was in linear wave equations \cite{unru81}) is addressed for future work. 

\section*{Acknowledgement}

{The research of FA and BG are supported by Riset Inovasi ITB 2016.} HS is grateful to the British Council for the 2015 Indonesia Second City Partnership Travel Grant.

\end{document}